# Deformation and failure mechanisms of Ti-6Al-4V as built by selective laser melting

Atieh Moridi [a,b,c*], Ali Gökhan Demir [c], Leonardo Caprio[c],

A. John Hart [a], Barbara Previtali [c], Bianca M. Colosimo [c]

[a] Department of Mechanical Engineering and Laboratory for Manufacturing and Productivity
Massachusetts Institute of Technology, Cambridge, MA, 02139, United States

[b] Sibley School of Mechanical and Aerospace Engineering, Cornell University, Ithaca, NY, 14853, United States

[c] Department of Mechanical Engineering, Politecnico di Milano, Milan, 20156, Italy

*corresponding author: moridi@cornell.edu

The ability to create complex geometries with tailored material properties has brought interest in using additive manufacturing (AM) techniques for many industrial applications. However, the complex relationship between AM process parameters, microstructure, and resultant properties of metals needs to be fully understood for the widespread use of metal AM. In this study, selective laser melting (SLM) is used to print Ti-6Al-4V. In-situ tensile tests with concurrent detailed microstructural analysis using electron backscatter diffraction, electron channeling contrast imaging, and digital image correlation are performed to understand the damage mechanism and its relation to the microstructure of Ti-6Al-4V after SLM processing. Our results demonstrate that the as-printed part shows hierarchical microstructures, consisting of primary, secondary, and tertiary $\alpha'$ martensite. This hierarchical structure is formed as a result of cyclic heat treatment during the layer-wise SLM process. Upon tensile deformation, strain localization within primary $\alpha'$ martensite results in microscopic ductile micro-void formation and coalescence, as well as macroscopic brittle fracture. In addition to localization inside primary $\alpha'$, surface steps at the boundaries of these high aspect ratio grains are formed which reveal the contribution of interfacial plasticity to the overall deformation of the material.

**Keywords:** Selective laser melting, additive manufacturing, Ti-6Al-4V, deformation, fracture, microstructure.

# 1 Introduction

High strength to weight ratio and good corrosion resistance make titanium alloys attractive for lightweight components, especially in aerospace and biomedical applications. For most structural applications, a two phase titanium alloy, consisting of a low temperature $\alpha$ phase and high temperature $\beta$ phase is used. The $\alpha + \beta$ titanium alloys can have different morphologies (i.e. fully lamellar, fully equiaxed, or a combination of the two) depending on the processing thermal history [1]. However, a fully lamellar structure is generally preferred for high fracture toughness and fatigue crack propagation resistance [2].

Additive manufacturing (AM) of titanium alloys is gaining an increasing attention because many titanium alloys are expensive and difficult to form/machine with conventional methods [3,4]. There are several methods for 3D printing metals including selective laser melting (SLM), electron beam melting (EBM), directed energy deposition (DED), and binder jetting [5,6]. In the present study, we focus on selective laser melting (SLM) of Ti-6Al-4V which typically exhibits a lamellar microstructure [7–15]. In SLM, a thin layer of metal powder is spread on a powder bed and selectively fused using a high power laser. This process is repeated layer by layer until the whole part is printed.

One of the key challenges associated with the adoption of AM technologies for the production of structural components is the uncertainties associated with the resultant mechanical properties. For example, the reported values of ductility for as-printed selective laser melted Ti-6Al-4V, spans from 1.7% up to 11.9% [16]. Such large scatter in the mechanical properties, in our view, is attributed to the microstructural heterogeneities that could arise during SLM.

Under the rapid solidification and cooling rates of SLM, high temperature β phase of Ti-6Al-4V alloy transforms into metastable hcp α′ martensite. Due to the high cooling rates, high density of defects such as dislocations and twins are generally present in the as-printed structures [17]. Printed layers are then subjected to multiple heating cycles with a decay peak temperature as subsequent layers are being deposited, which in turn could modify the microstructure [15]. Epitaxial growth from the previous layer upon successive laser scans can result in columnar β grains [18,19]. Preferential growth orientation of $α′$ martensite is aligned at about 40° with respect to the build direction [10,13]. This orientation is pretty close to the maximum macroscopic shear stress if tensile loading normal to the build direction is applied. Existence of thin films of retained β phase at the interface between martensites has also been reported [15,20]. Many studies in the literature have focused on mechanical properties of selective laser melted Ti-6Al-4V [19,21,22]. The majority of data in the literature attribute the brittle fracture of selective laser melted Ti-6Al-4V to two main origins: (i) the $α′$ martensitic phase and (ii) the defects that are generally present in the printed structure. Different strategies have been accordingly applied to enhance the ductility of printed Ti-6Al-4V. To address the former, in situ decomposition of brittle $α′$ phase into more ductile α + β phases by tuning laser parameters is proposed [19]. To address the latter, optimization of process parameters and developing processing maps to minimize defects have been proposed [21,22].

Recently, it has been argued that the tensile yield strength and uniform elongation of printed Ti-6Al-4V is mainly governed by the as-built microstructure, while the strain-to-failure is sensitive to the porosity, even in very high-density samples [17]. However, what microstructural features govern the deformation is not thoroughly investigated. In addition, the role of martensite interface on the overall deformation of the material has not been studied in the past. What is more, brittle and ductile features are typically observed on the fracture surface of Ti-6Al-4V processed by SLM [11,21,23]. However, the mechanism leading to this pseudo-brittle failure is not clear yet.

The aim of this work is to understand the deformation mechanisms and damage in selective laser melted Ti-6Al-4V. In general, the ductility of printed Ti-6Al-4V in the transverse direction is lower than the build direction [16,24]. Therefore, the focus of this work is on the properties in the transverse direction. In-situ tensile tests with detailed microstructural analysis using electron backscatter diffraction (EBSD), electron channeling contrast imaging (ECCI), and digital image correlation (DIC) are performed to establish the connection between microstructure and mechanical properties.

## 2   Method

An open platform SLM has been used for printing Ti-6Al-4V powder under Argon atmosphere [25]. Printing process parameters were chosen based on prior porosity optimization experiments. The laser power and scan velocity were set at 300W and 300 mm/s respectively. The layer height and hatch spacing were fixed at 50 microns. Ti-6Al-4V powder (LPW Technology Ltd, UK) with nominal size distribution between 10-45 μm was used. Titanuim plates with 10 mm thickness were used as a build platform. The scan path trajectory was designed such that the entire gauge length of tensile samples was printed in one island. For each subsequent layer, the island pattern was rotated by 45° as shown in Fig. 1a.

Tensile samples were cut from rectangular builds by electrical discharge machining (EDM). The gauge cross-section was 2mm (width) × 0.75mm (thickness) × 6 mm (gage length). Build direction (BD) is perpendicular to the tensile axis. Tensile samples were grinded and polished for surface observations. Tensile tests were performed at room temperature and nominal strain rate of $1 \times 10^{-4}\,\text{s}^{-1}$ using Gatan Microtest module with 2 kN load capacity. Three samples were tested to analyze the repeatability of the results.

Nanoindentation was performed using a Hysitron triboindenter and a Berkovich tip. All indents were quasi-static with 30s loading up to 250 μN, 5 s hold, and 30 s unloading.

The macroscale strain distribution during tensile testing was observed by optical DIC based on spray coating. Secondary electron (SE) and back scatter electron (BSE) imaging and EBSD was carried out using a TESCAN Mira scanning electron microscope (SEM) operated at 30 KV and equipped with low energy back scatter detector and EDAX-EBSD camera. The high resolution DIC (HR-DIC) was performed by secondary electron (SE) imaging of the $SiO_2$ particles on the surface [26]. SE imaging on the same microstructural position was repeated at 0.15% strain steps. The SE images were correlated using digital image correlation (DIC) in the GOM software. Before deformation, EBSD was carried out on the same microstructural positions as observed with SE imaging during deformation. All EBSD measurements were performed with step size of 50 nm. Samples for EBSD were prepared by mechanical polishing using several SiC sand papers and diamond suspensions up to 1 μm, followed by 0.5 hour polishing using a $SiO_2$ colloidal suspension. Reconstruction of the prior $\beta$ grain structures is achieved by the ARPGE software [27]. For automated analysis of highest Schmid factor slip systems, Orientation Imaging Microscopy (OIM) Analysis™ provided by EDAX is used. For large field of view imaging, specimens were chemically etched using Keller etchant.

# 3 Results:

## 3.1 Mechanical properties and pseudo-embrittlement:

In Fig. 1b we show the macroscopic tensile properties of the printed Ti-6Al-4V in the transverse direction. The ultimate tensile strength is 1177±36 MPa and the overall ductility is 2.0±0.05 %. The strength and ductility of the printed parts show good repeatability revealed by small error bars on the plot. Fig. 1c shows the DIC map of deformation where the local strain near the fracture zone goes up to 4.5%. Our measured tensile strength agrees well with values reported in the literature ranging from 1006 to 1327 MPa [16]. The strain to failure reported in the literature, on the other hand, widely varies from 1.6 to 11.9% [16]. In comparing the reported range with our measurement of ductility one should note that most of the deformation in more ductile samples happens post-necking and the uniform elongation is generally limited to 2-4% [17]. This latter range agrees well with our measurements.

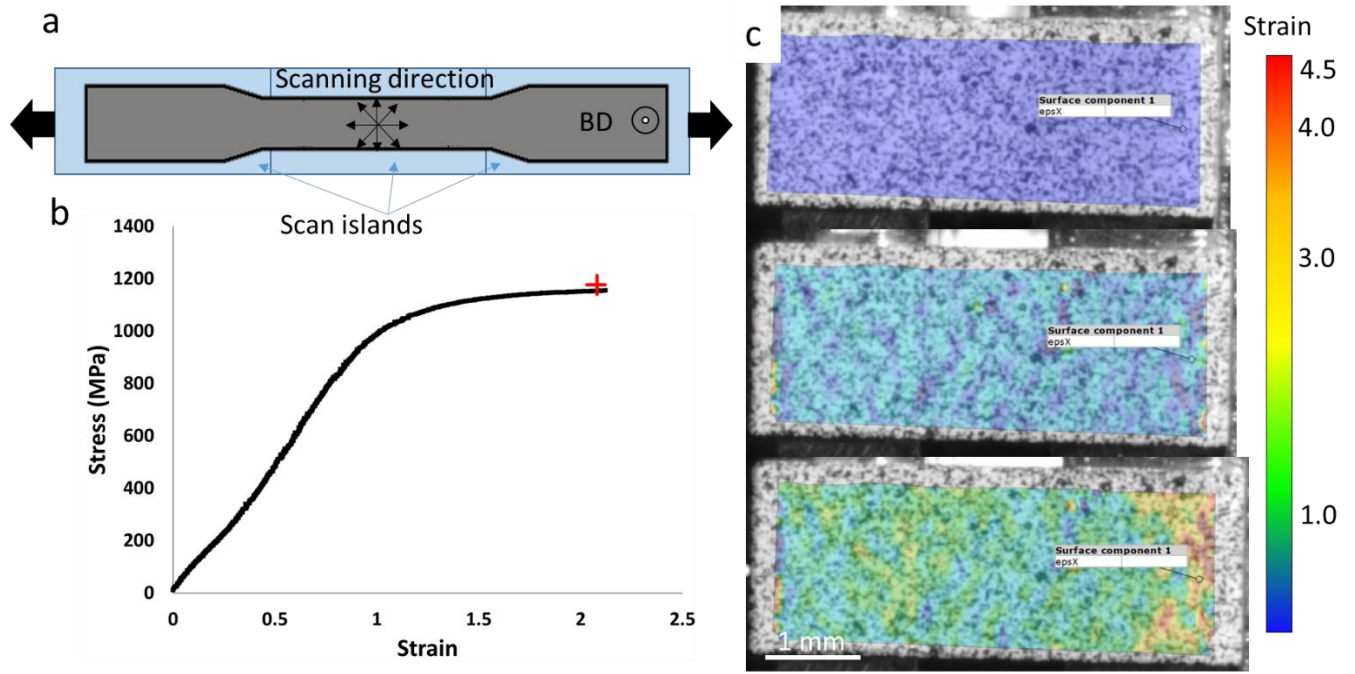

*Figure 1. Sample geometry and mechanical behavior a) Build geometry and planar scan strategy where for each subsequent layer, the island pattern was rotated by 45° b) engineering stress-strain behavior of as-printed Ti-6Al-4V in transverse direction, c) local strain evolution in the gauge length during deformation (undeformed, 1 and 2% strain).*

Fig. 2a shows the macroscale fracture surface of one of the broken samples. The overview of fracture surface shows planar, faceted features that are highlighted in the image. The detailed view of these facets (Fig. 2b) show a dominant cleavage-type fracture surface surrounded by dimple like features (see the arrows). While the former represents a brittle fracture, the latter is a characteristic of ductile fracture. The post fracture cross section of the sample (cut in the mid plane in the longitudinal direction) is shown in Fig. 2c. It is interesting to note that this cross section consists of microvoids oriented 45° with respect to the loading direction (LD), corresponding to the plane of maximum shear stress at the macroscale. The fact that the fracture surface follows the same lines suggests that these microvoid nucleation and coalescence resulted in the final fracture.

Microvoid formation is a characteristic of ductile fracture whereas the resulting apparent macroscopic behavior of the printed samples is brittle. These voids are not pre-existing voids due to the processing such as lack of fusion or gas entrapment defects because they are not present far from the fracture surface (beyond the horizental dashed line in Fig. 2c). Therefore, these voids are formed during the deformation of the material causing pseudo-embrittlement. The term pseudo-embrittlement refers to promotion of a local ductile fracture resulting in deterioration of the global ductility [28,29]. To further understand the origin of microvoid formation, detailed microstructural analysis is performed which will be presented in the following sections.

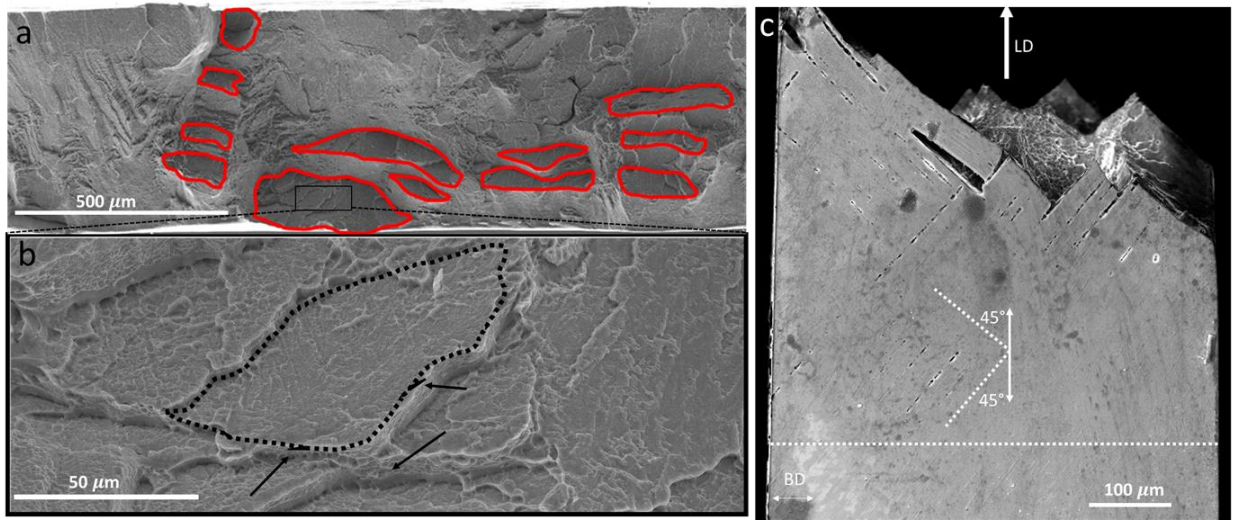

*Figure 2. Fracture surface of as-printed Ti-6Al-4V a) overview of the fracture surface b) detailed view of the fracture surface showing cleavage facets (dashed line) surrounded by dimples (arrows). c) cross section in longitudinal direction of sample after fracture showing voids that are oriented 45° with respect to the tensile axis.*

### 3.2 Heterogeneous microstructure of as-printed Ti-6Al-4V

The microstructure of printed Ti-6Al-4V in the transverse direction is shown in Fig. 3a. The thermal gradient in the build direction leads to coarse columnar β-grain structures as shown by the dashed line in the Fig. 3a. During solidification, the high temperature β-phase transforms to metastable α′ martensite. However, the transformation path is complicated since the solidified material then experiences several thermal cycles, even above the β-transus temperature, from subsequent beam passes as shown schematically in the inset in Fig. 3a. Complex thermal history results in formation of substructures in the solidified material. The detailed view of the boxed region in Fig. 3a shows different grain morphologies that are present in the microstrructure. The primary α′ grains are transformed early on during printing and are extended throughout the beta grain boundary. These grains are also subjected to autotempering during the cyclic heat treatment and thus are less dislocated and softer than the surrounding microstructure. Secondary and tertiary α′ grains are transformed later on during the process and their growth is inhibited by the boundaries of previously formed grains. The histogram of different types of α′ grains in three images (each 50×50 $\mu m^2$) is shown in Fig. 3c. There are only a few primary α′ grains with areas between 10-50 $\mu m^2$ and the quantity enhances as we move toward smaller grains. A certain amount of β stabilizer (V) in Ti-6Al-4V alloy can promote the coexistence of β and α′. The β phase is revealed by the brighter contrast in the etched microstructure in Fig. 3b. Presence of β phase in the as-printed microstucture has also been reported in the literature [15,20].

To verify the difference in hardness between the primary α′ and the surrounding substructure, we conducted 100 nanoindentations on a 20 by 20 $\mu m^2$ grid. The hardness values presented in Fig. 3d shows a large scatter spanning from 4.3 GPa to 7.12 GPa. Arrows show the hardness values of indentations inside primary α′ that are all in the first quarter of the hardness range. It should be noted that variation in the hardness values are both due to size effect (i.e. primary martensite is larger than secondary and tertiary martensites) as well as thermal history effect (i.e. primary martensite is subjected to more autotempering).

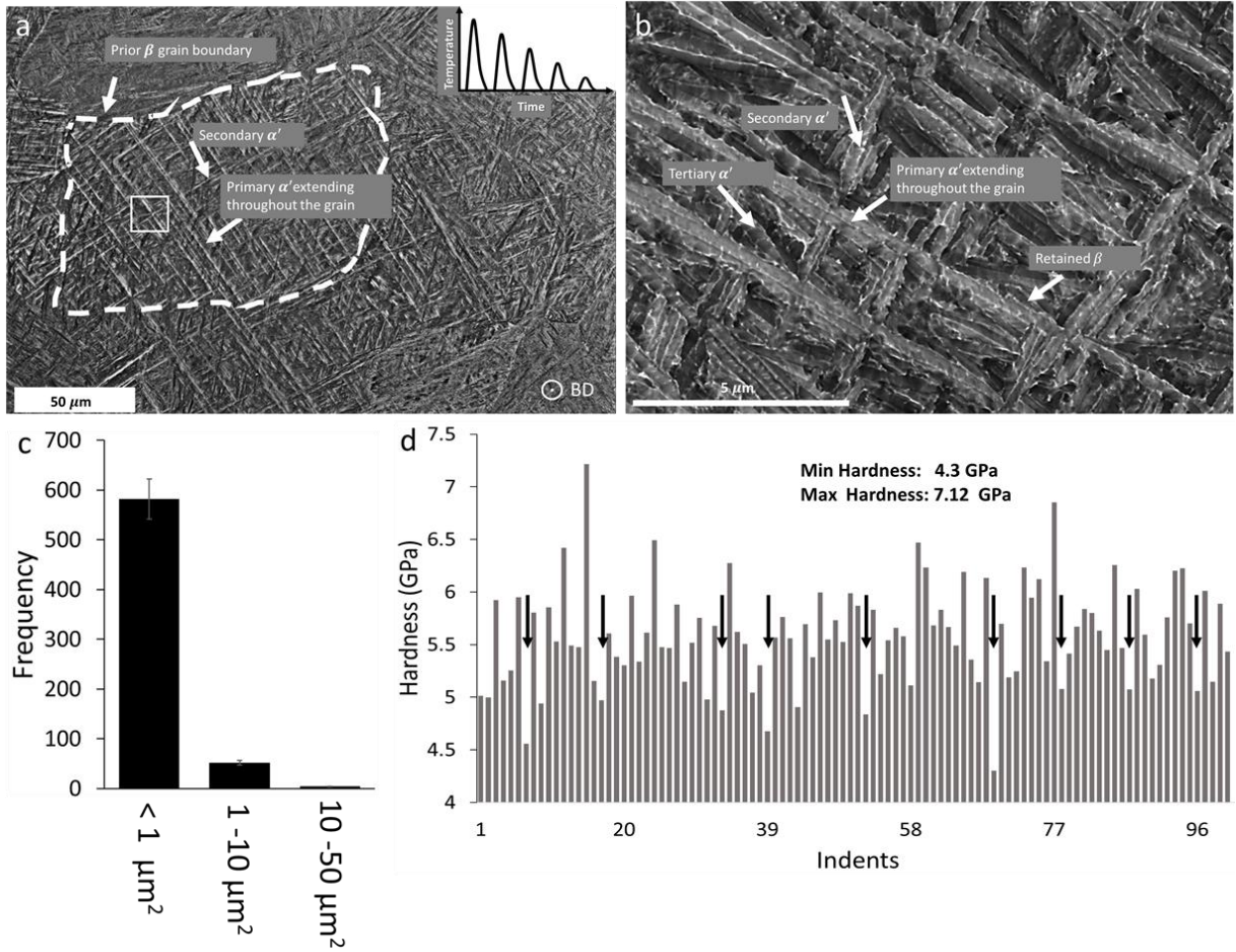

*Figure 3. Heterogeneous microstructure a) The microstructure of printed Ti-6Al-4V in transverse direction, the dashed line shows prior beta grain boundary, the inset shows the schematic of thermal history resulting in this heterogeneous microstructure b) detailed view of the box in panel a highlighting different α' morphologies and retained beta phase. c) frequency of different α' morphologies in three images of 50×50 μm² and d) hardness distribution in the heterogeneous microstructure, arrows show hardness values of indentations inside primary α'.*

## 3.3 Strain localization

To have a better understanding of the relationship between heterogeneous microstructure and deformation, in situ tensile testing was performed. We first analyzed the axial micro-strain distribution during the deformation of the as-printed Ti-6Al-4V. Three steps are shown in Fig. 4 a-c corresponding to the undeformed, global strain of 1%, and 2% respectively. Individual strain bands in Fig. 4c are predominantly oriented at 45° with respect to the loading direction (corresponding to the highest shear strain at the macroscale).

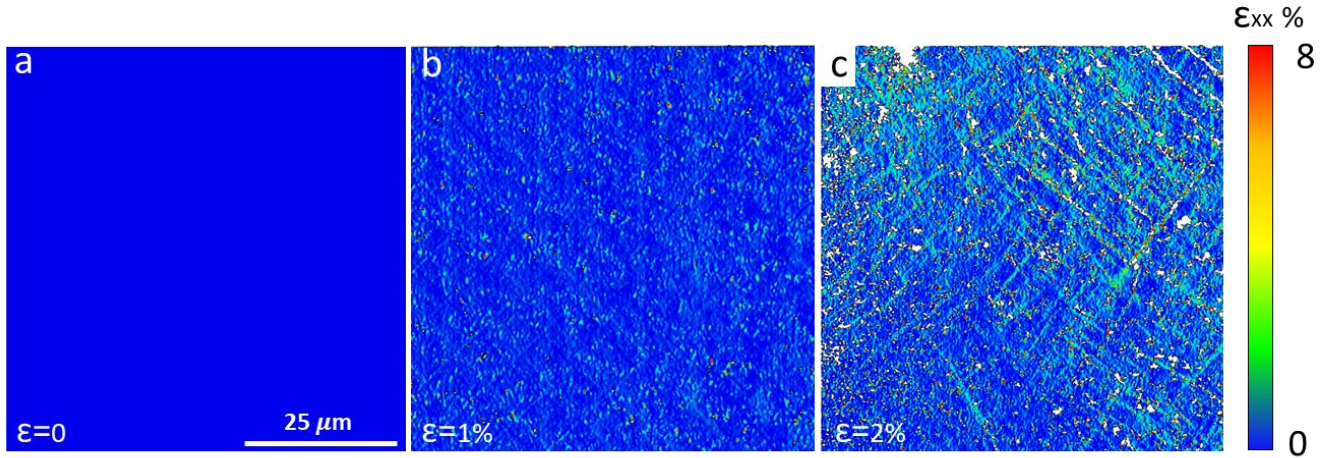

Figure 4. micro-strain map in a) undeformed b) 1% and c) 2% global strain states.

In Fig. 5 a and b, we correlated the HR-DIC map at 2% global strain with the EBSD grain size map of the same region. We could identify two features in this correlation. First, it is apparent that the grains with high aspect ratios (and therefore bigger in size), that are oriented at approximately 45° to the loading direction, correspond well with the localization bands in the HR-DIC map. Several of these grains are highlighted with the solid ellipsoids in Fig. 5 a and b. This is in good agreement with our hardness measurements where the high aspect ratio laths represented soft spots in the microstructure. Second, the colony of laths, where all laths belong to the same crystallographic variant (as shown in the inverse pole figure (IPF) map in Fig. 5c), also show a high strain localization tendency as highlighted by dashed ellipsoids. Both colony structures are favorably oriented parallel to the plane of maximum macroscale shear direction.

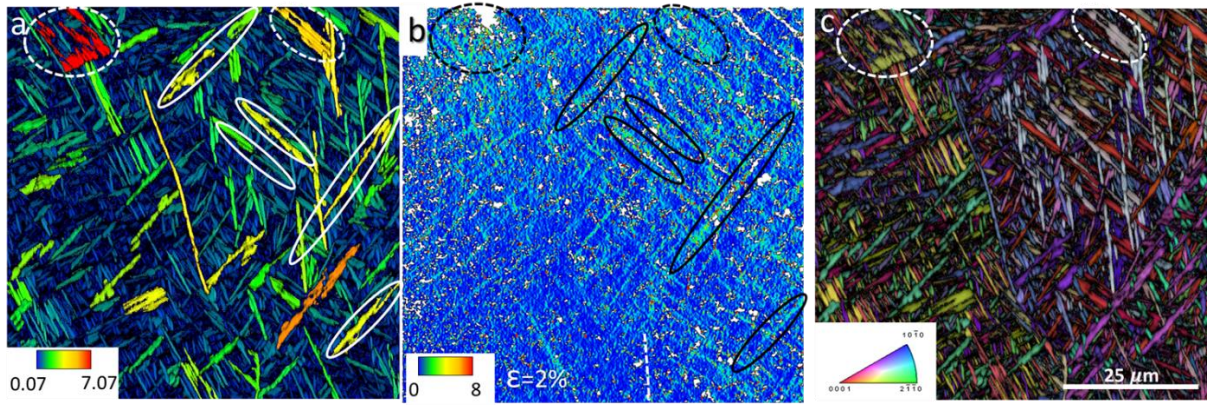

*Figure 5. Effect of microstructure on strain localization a) EBSD grain size map b) micro-strain map at 2% global strain showing localization tendency at high aspect ratio α′ laths that are oriented at 45° to the loading axis (highlighted by solid ellipsoids) and α′ colonies highlighted by dashed ellipsoid and c) EBSD Inverse Pole Figure (IPF) map.*

### 3.4 Deformation mechanism

Plastic deformation in Ti-6Al-4V is mostly accommodated by basal or prismatic slip [1]. The overlay of image quality map and Schmid factor map for the basal and prismatic slip systems are shown in Fig. 6a and b respectively (same field of view as Fig. 4 and 5). Only grains with high Schmid factor (above 0.4) are colored in the figure for clarity. Due to the specific orientation relationship between β and α′ phase ($[0001]α // [110]β$ and $[11\bar{2}0]α // [111]β$), the parent β grains can be reconstructed based on the local texture [27] as shown in Fig. 6c . The surface topography of the sample after deformation is shown in Fig. 6d. The change in contrast in this figure shows good agreement with the deformation localization map in Fig. 4c (i.e. higher strain corresponds to brighter contrast). Surface topography evolution has been shown to be an effective method for local strain mapping [30].

Comparing the Schmid factor maps with the HR-DIC map in Fig 4c or surface topography in Fig 6d demonstrates that strain localization in the high aspect ratio α′ laths occurs primarily in grains with high Schmid factors in the prismatic slip system (shown by the arrows in Fig. 6b and d). In particular, inside grain 2 with high density of martensites with high basal Schmid factor, strain localization occurs in a few grains with high prismatic Schmid factor. On the other hand, both $\alpha'$ colonies with strain localization in HR-DIC map of Fig. 5, show high basal Schmidt factor as highlighted by ellipsoids.

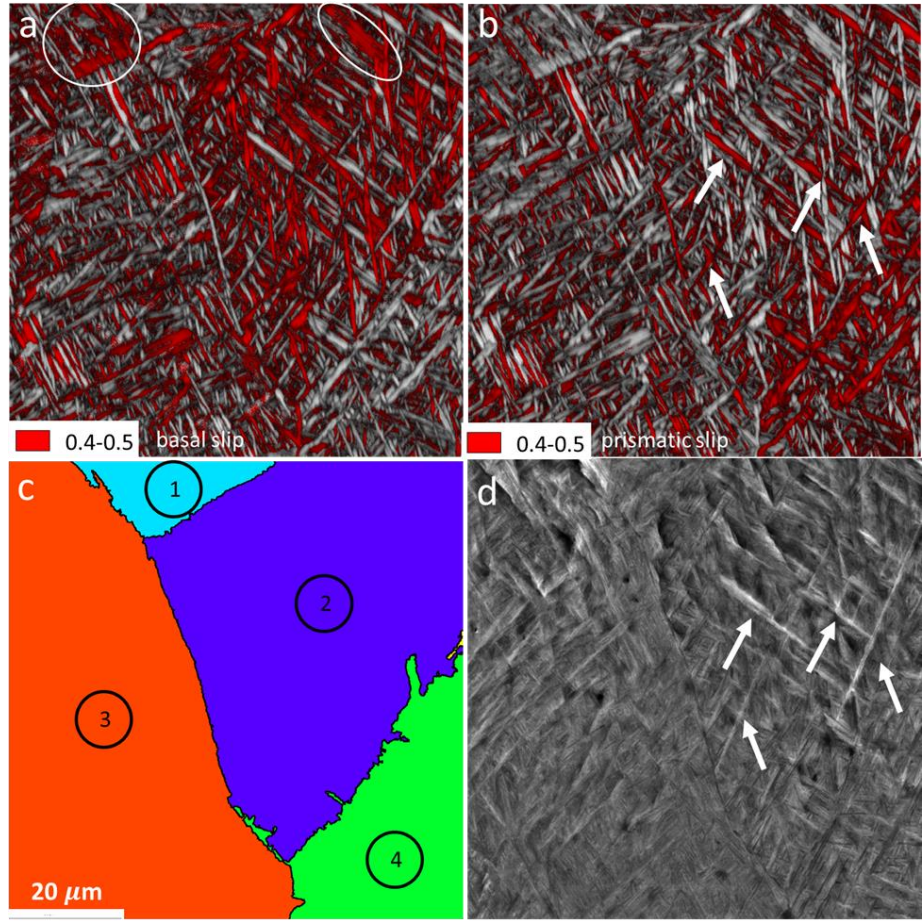

*Figure 6. Effect of schmid factor on strain localization a) Schmid factor map for the basal and b) prismatic slip systems; red color shows grains with schmid factor between 0.4-0.5, c) reconstructed prior beta grain boundary of the same region, d) surface topography after deformation.*

Based on the surface topography evolution (and its correspondence with the HR-DIC map), we can study the large scale/bulk plasticity of the printed Ti-6Al-4V. Fig. 7a shows a wide field of view of the deformed sample close to fracture surface. Loading is applied in the horizontal direction. It is apparent that the steep surface steps are confined within the prior beta grain boundaries. This implies that reducing the beta grain size could be an effective method in improving the mechanical properties of the printed Ti-6Al-4V parts. The majority of surface steps are oriented at approximately 45° to the loading direction. However, there are a few grains (for example, grains 5 and 6) where the $\alpha'$'s major axis is not aligned with the maximum shear stress direction and yet these grains undergo localization. The surface steps in grains 5 and 6 are almost perpendicular to the loading direction. In the majority of the grains, there are steep surface steps only in one major direction, but there are also a few grains (for example, grains 3 and 4) where localization occurs in two perpendicular directions implying that localization can happen in the secondary $\alpha'$ as well. The detailed view of surface topography shows that there are two distinct scales of surface steps as shown in Fig. 7b suggesting that two different deformation mechanisms must be activated. The surface steps are identified by contrast changes as shown in the inset in Fig. 7b (the plot shows gray values across the dashed line in Fig. 7b). The smaller steps (gray arrows in Fig. 7b) are slip traces associated with dislocation activity which result from the intersection of an active slip plane with the sample surface. Fig. 7c shows ECCI of the deformed microstructure showing dislocation network inside the grains. The bigger steps, on the other hand, are at the grain boundaries (Fig. 7d), indicating the contribution of interface plasticity to the overall deformation of the material. Grain boundary is accommodating the plastic deformation incompatibility between the adjacent grains without cracking which could be attributed to the thin film of retained $\beta$ phase at the boundaries acting as a lubricant.

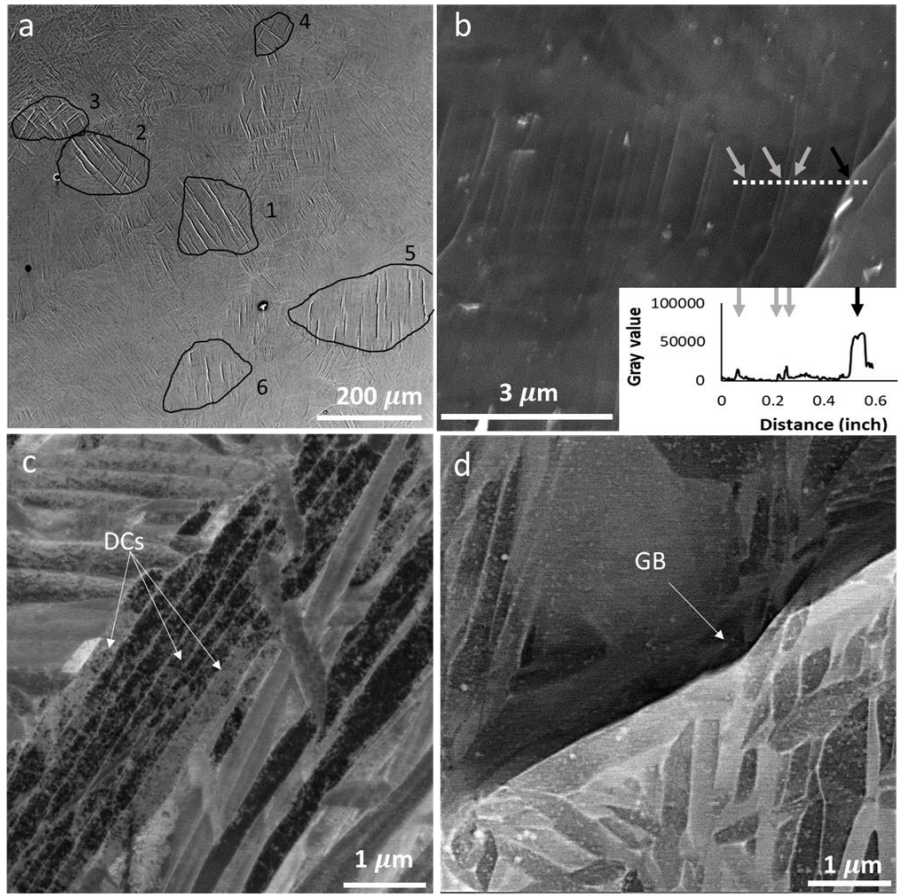

*Figure 7. Surface topography during deformation a) overview of deformed sample close to fracture surface showing steep surface steps confined within prior beta boundaries, b) two scale of surface step associated with c) dislocation activities and d) boundary step showing interfacial plasticity.*

## 3.5 Damage

Fig. 8a is a cropped view of the fracture cross section presented earlier in Fig. 2c. A detailed view (Fig. 8b) shows that these voids are mainly inside the primary α′. We can also observe indications of other localized plasticity such as rippling effect at the boundaries of α′ which can potentially lead to void formation (Fig. 8c) or necking and rupture of the primary α′ (Fig. 8d) in the deformed microstructure. These observations suggest that the local damage is promoted by a highly ductile process, and mostly confined within primary α′ grains. During deformation, localized, microscopically ductile microvoid formation and coalescence along primary $α′$ martensite result in macroscopic brittle fracture. Localization of plastic deformation during early stages of loading reduces the contribution of the whole cross section to bear the applied load. This can also explain the ductile and brittle nature of the fracture surface which is commonly observed in AM parts and presented in Fig. 2a and b. These observations imply that localization in high aspect ratio grains leading to void nucleation and coalescence is the predominant cause of failure in the as-printed Ti-6Al-4V.

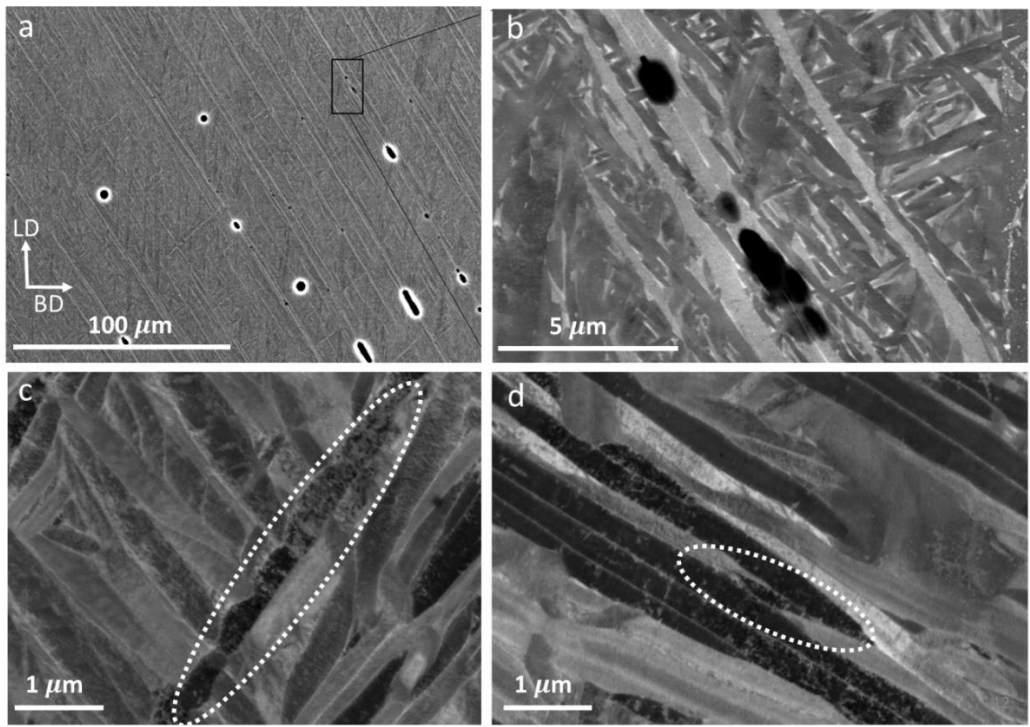

*Figure 8. Ductile local damage a) overview of the cross section of broken sample with voids aligned 45° to the loading direction b) void are located inside primary α′ c) rippling effect at the boundaries of primary α′ and d) necking and rupture of $primary\ α′$.*

## 4 Discussion

Based on the results presented in the previous sections, it is now possible to discuss the mechanism of deformation and damage in as-built additively manufactured Ti-6Al-4V. Under the rapid solidification and cooling rates of SLM, BCC β phase of Ti-6Al-4V alloy transforms completely into metastable hcp α′ martensite phase by a diffusionless, shear-type transformation process. The shear deformation involved in martensitic transformation results in a lath structure with high initial dislocation density. The cyclic heat treatment during AM can lead to some degree of autotempering in the solidified material. The degree of autotempering is mainly dependent on the sequence of martensitic transformation. High aspect ratio laths, that are transformed early on during AM process and span throughout the prior β grain boundary, are subjected to more autotempering and are less dislocated. The boundaries of these high aspect ratio grains impede growth of laths that are transformed later in the process. In other words, the deformability is increasing from late to early martensite (due to more autotempering and therefore less dislocation) resulting in a heterogenous microstructure as schematically shown in Fig. 9a. Such heterogeneity based on the sequence of martensitic transformation has also been observed for martensitic steel [31].

Upon deformation, strain localizes primarily in high aspect ratio laths that represent soft spots in the microstructure as shown in Fig. 9b. Other than being less dislocated, local plasticity is enhanced in high aspect ratio laths by a long dislocation mean free path in the lath's longitudinal direction. We also observed strain localization in Colonies of $\alpha'$ lamellae as shown in Fig. 5. However, the large field-of-view of deformed microstructure with steep surface steps (shown in Fig. 7a.) suggest that localization in high aspect ratio laths is more predominant.

During deformation, $\alpha'$ major axis orientation and slip system orientation can affect the deformation response. Localization occurs more frequently in the laths that are oriented 45° to the loading direction. High prismatic and basal Schmid factors also show good correlation with strain localization in the high aspect ratio $\alpha'$ and $\alpha'$ colonies respectively. However, correlation between strain localization in high aspect ratio $\alpha'$ and high basal Schmid factor has also been reported in the literature [14]. At room temperature, the prismatic slip system in the α-Titanium displays the lowest critical resolved shear stress [32]. Increasing Al content in the alloy tends to reduce the ratio of critical resolved shear stress (CRSS) values for basal and prismatic slip. The stress to activate these two systems becomes equal at 5.74 wt% Al [32,33]. This could explain why the correlation between strain localization and both high basal and prismatic Schmid factor can be observed. It is also important to note that the Schmid factor maps are obtained by projecting the applied macroscopic stress onto the slip systems of each grain based on their orientation. However, strain incompatibilities between grains and elastic and plastic anisotropies resulting from grain interactions result in complex stress states in each grain. Therefore, it has been argued that the stress tensor in each grain as opposed to the macroscopic applied stress should be used to determine the active slip systems [34]. It has also been shown that there is not a correlation between the Schmid factor and high strain values measured by DIC in FCC materials [35]. On the other hand, Schmid law has been shown to be satisfactory to explain slip system activation in Ti-6Al-4V [14,36] and good correlation in the present study is also observed.

Surface topography studies (Fig. 7a and d) show that plasticity is taking place at the martensite boundaries resulting in boundary surface steps. The effect of interface plasticity on the deformation of printed Ti-6Al-4V has not been studied so far. As high aspect ratio laths do not have any internal boundaries, their grain boundaries are especially prone to interface plasticity to accommodate the incompatibilities with the surrounding microstructure. The boundary can accommodate significant out of plane deformation without cracking due to the presence of thin beta films at the interface acting as lubricant. Strain localization in primary $\alpha'$ is confined within the prior $\beta$ boundaries. No clear localization at prior $\beta$ boundaries is observed in the large field-of view image in Fig. 7a. Fig. 9b schematically represent two major deformation mechanisms that are slip and interfacial plasticity.

Strain localization in the primary $\alpha'$ leads to void nucleation as schematically presented in Fig. 9c. The voids coalesce causes the final fracture. A previous synchrotron X-ray computed tomography (SXCT) experiment has also detected pore formation inside AM Ti-6Al-4V after deformation, albeit at much lower resolution [17]. We argue that these pores are not pre-existing voids due to processing (as these pores do not exist far from the fracture zone), but rather are generated due to the localized plasticity during deformation.

The localized plasticity, promoted as a result of the heterogeneous microstructure, is a highly ductile process but results in a macroscale brittle fracture. The fracture surface (shown in Fig. 2) also has characteristics of both ductile and brittle failure. This is the reason why we use the term "pseudo embrittlement" to refer to this localized ductile process deteriorating the overall ductility of the material. Void nucleation in the material, reduces the overall solid cross section, and results in reduction of the macroscopic load in the load-displacement data. In our view, this can be the reason why all selective laser melted Ti-6Al-4V have limited uniform elongation (between 2-4%) [17] and show early necking irrespective of their overall ductility (which varies from 1.6 to 11.9% [16]). Up to the necking point, the behavior of the material is mainly dependent on the heterogeneous microstructure. Beyond the necking point, ductility can be dependent on the sample geometry [37]. It is important to note that the gage section of our samples (2mm×0.75mm) for micromechanical testing is an order of magnitude smaller than most of the tensile samples used in the literature which are normally cylindrical tensile specimens with diameter between 5 and 10 mm [21,23]. Our samples do not show a clear necking before fracture. Tensile specimens with a bigger gauge cross section on the other hand, tend to show higher elongation to failure presumably due to bigger cross section for growth and coalescence of the voids.

We believe the findings of this work can help develop strategies toward printing Ti-6Al-4V microstructures with enhanced mechanical properties. Improving the homogeneity of the microstructure through in-situ scanning strategies or refining the beta grain size by modifying the chemistry are examples of such developments. Understanding the effect of post heat treatment or hot isostatic pressing on microstructure and mechanical properties of printed Ti-6Al-4V was beyond the scope of this paper but is the subject of an ongoing study.

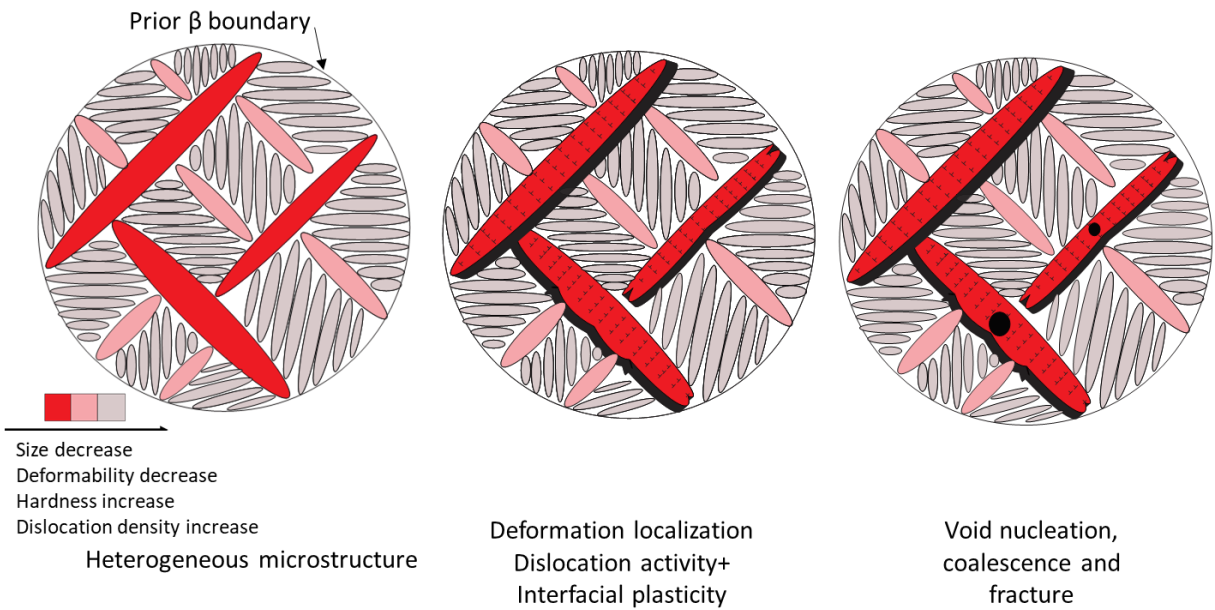

*Figure 9. Schematic representation of deformation and damage in additively manufactured Ti-6Al-4V a) heterogeneous microstructure where deformability decreases from early to late transformed martensite b) deformation localization in high aspect ratio laths that also leads to interfacial plasticity (only dislocations generated during deformation are shown) and c) deformation localization leads to void nucleation, the voids grow and coalesce and result in final fracture.*

## 5 Conclusion

We investigated the deformation response and damage mechanism of selective laser melted Ti-6Al-4V using an in situ tensile test and post-mortem fracture analysis. Based on the results the following conclusions are drawn:

- Selective laser melting results in a hierarchical microstructure based on the sequence of martensitic transformation. Early martensites (i.e. primary $\alpha'$) span throughout the prior $\beta$ grain boundary and are subjected to more autotemperaing, thus less dislocated and softer. Late martensites, are smaller because their growth is stopped by the boundaries of early martensites, subjected to less autotempering and therefore have more dislocations and are harder.
- Upon tensile deformation, the strain mainly localizes in primary $\alpha'$. Localization mainly happens in the primary $\alpha'$ with major axis oriented at approximately 45° with respect to the loading direction.
- Evolution of the surface topography during deformation reveals the role of interface plasticity on the overall deformation of the printed Ti-6Al-4V. The boundaries of the primary $\alpha'$ are particularly prone to interface plasticity to accommodate strain incompatibilities with the surrounding microstructure. No significant localization at the prior $\beta$ grain boundaries is observed
- Strain localization results in the void formation within the primary $\alpha'$. Void growth and coalescence result in the final fracture of the material. Local damage is highly ductile but results in an apparent brittle failure. This pseudo-embrittlement is also seen in the fracture surfaces which contain both dimple and cleavage features.

# 6 Acknowledgment

A.M. acknowledges financial support from the Polimi International Fellowship (PIF). A. M, B. M. C. and A. J. H acknowledge financial support from the MISTI global seed fund. All authors acknowledge Prof. Cem Tasan for discussion and for allowing us to use the in situ mechanical testing facility in his lab.# 7 References

[1]   C. Leyens, M. Peters, Titanium an Titanium Alloys, 2003. doi:10.1002/3527602119.

[2]   Z. Tarzimoghadam, S. Sandlöbes, K.G. Pradeep, D. Raabe, Microstructure design and mechanical properties in a near-α Ti-4Mo alloy, Acta Mater. 97 (2015) 291–304. doi:10.1016/j.actamat.2015.06.043.

[3]   M. Nouari, H. Makich, On the Physics of Machining Titanium Alloys: Interactions between Cutting Parameters, Microstructure and Tool Wear, Metals (Basel). 4 (2014) 335–358. doi:10.3390/met4030335.

[4]   M. Nouari, H. Makich, Experimental investigation on the effect of the material microstructure on tool wear when machining hard titanium alloys: Ti-6Al-4V and Ti-555, Int. J. Refract. Met. Hard Mater. 41 (2013) 259–269. doi:10.1016/j.ijrmhm.2013.04.011.

[5]   S. Gorsse, C. Hutchinson, M. Gouné, R. Banerjee, Additive manufacturing of metals: a brief review of the characteristic microstructures and properties of steels, Ti-6Al-4V and high-entropy alloys, Sci. Technol. Adv. Mater. 18 (2017) 584–610. doi:10.1080/14686996.2017.1361305.

[6]   W.J. Sames, F.A. List, S. Pannala, R.R. Dehoff, S.S. Babu, The metallurgy and processing science of metal additive manufacturing, Int. Mater. Rev. 61 (2016) 315–360. doi:10.1080/09506608.2015.1116649.

[7]   L. Liu, M. He, X. Xu, C. Zhao, Y. Gan, J. Lin, J. Luo, J. Lin, Preliminary study on the corrosion resistance, antibacterial activity and cytotoxicity of selective-laser-melted Ti6Al4V-xCu alloys, Mater. Sci. Eng. C. 72 (2017) 631–640. doi:10.1016/j.msec.2016.11.126.

[8]   B. Song, S. Dong, B. Zhang, H. Liao, C. Coddet, Effects of processing parameters on microstructure and mechanical property of selective laser melted Ti6Al4V, Mater. Des. 35 (2012) 120–125. doi:10.1016/j.matdes.2011.09.051.

[9]   F. Li, Z. Wang, X. Zeng, Microstructures and mechanical properties of Ti6Al4V alloy fabricated by multi-laser beam selective laser melting, Mater. Lett. 199 (2017) 79–83. doi:10.1016/j.matlet.2017.04.050.

[10]  C. Qiu, N.J.E. Adkins, M.M. Attallah, Microstructure and tensile properties of selectively laser-melted and of HIPed laser-melted Ti-6Al-4V, Mater. Sci. Eng. A. 578 (2013) 230–239. doi:10.1016/j.msea.2013.04.099.

[11]  L. Facchini, E. Magalini, P. Robotti, A. Molinari, S. Höges, K. Wissenbach, Ductility of a
22

Ti-6Al-4V alloy produced by selective laser melting of prealloyed powders, Rapid Prototyp. J. 16 (2010) 450–459. doi:10.1108/13552541011083371.

[12] H.K. Rafi, T.L. Starr, B.E. Stucker, A comparison of the tensile, fatigue, and fracture behavior of Ti-6Al-4V and 15-5 PH stainless steel parts made by selective laser melting, Int. J. Adv. Manuf. Technol. 69 (2013) 1299–1309. doi:10.1007/s00170-013-5106-7.

[13] B. Vrancken, L. Thijs, J. Kruth, J. Van Humbeeck, J. Van Humbeeck, Heat treatment of Ti6Al4V produced by Selective Laser Melting : Microstructure and Mechanical properties, J. Alloys Compd. 541 (2012) 177–185. doi:10.1016/j.jallcom.2012.07.022.

[14] T.A. Book, M.D. Sangid, Strain localization in Ti-6Al-4V Widmanstätten microstructures produced by additive manufacturing, Mater. Charact. 122 (2016) 104–112. doi:10.1016/j.matchar.2016.10.018.

[15] J. Yang, H. Yu, J. Yin, M. Gao, Z. Wang, X. Zeng, Formation and control of martensite in Ti-6Al-4V alloy produced by selective laser melting, Mater. Des. 108 (2016) 308–318. doi:10.1016/j.matdes.2016.06.117.

[16] A.M. Beese, B.E. Carroll, Review of Mechanical Properties of Ti-6Al-4V Made by Laser-Based Additive Manufacturing Using Powder Feedstock, Jom. 68 (2016) 724–734. doi:10.1007/s11837-015-1759-z.

[17] T. Voisin, N.P. Calta, S.A. Khairallah, J.B. Forien, L. Balogh, R.W. Cunningham, A.D. Rollett, Y.M. Wang, Defects-dictated tensile properties of selective laser melted Ti-6Al-4V, Mater. Des. 158 (2018) 113–126. doi:10.1016/j.matdes.2018.08.004.

[18] M. Simonelli, Y.Y. Tse, C. Tuck, Microstructure of Ti-6Al-4V produced by selective laser melting, in: J. Phys. Conf. Ser., 2012. doi:10.1088/1742-6596/371/1/012084.

[19] W. Xu, E.W. Lui, A. Pateras, M. Qian, M. Brandt, In situ tailoring microstructure in additively manufactured Ti-6Al-4V for superior mechanical performance, Acta Mater. 125 (2017) 390–400. doi:10.1016/j.actamat.2016.12.027.

[20] H.Z. Zhong, M. Qian, W. Hou, X.Y. Zhang, J.F. Gu, The β phase evolution in Ti-6Al-4V additively manufactured by laser metal deposition due to cyclic phase transformations, Mater. Lett. 216 (2018) 50–53. doi:10.1016/j.matlet.2017.12.140.

[21] G. Kasperovich, J. Hausmann, Improvement of fatigue resistance and ductility of TiAl6V4 processed by selective laser melting, J. Mater. Process. Technol. 220 (2015) 202–214. doi:10.1016/j.jmatprotec.2015.01.025.

[22] M. Thomas, G.J. Baxter, I. Todd, Normalised model-based processing diagrams for additive layer manufacture of engineering alloys, Acta Mater. 108 (2016) 26–35. doi:10.1016/j.actamat.2016.02.025.

[23] W. Xu, M. Brandt, S. Sun, J. Elambasseril, Q. Liu, K. Latham, K. Xia, M. Qian, Additive manufacturing of strong and ductile Ti-6Al-4V by selective laser melting via in situ martensite decomposition, Acta Mater. 85 (2015) 74–84. doi:10.1016/j.actamat.2014.11.028.

[24] A.E. Wilson-Heid, Z. Wang, B. McCornac, A.M. Beese, Quantitative relationship

between anisotropic strain to failure and grain morphology in additively manufactured Ti-6Al-4V, Mater. Sci. Eng. A. 706 (2017) 287–294. doi:10.1016/j.msea.2017.09.017.

[25] A.G. Demir, L. Monguzzi, B. Previtali, Selective laser melting of pure Zn with high density for biodegradable implant manufacturing, Addit. Manuf. 15 (2017) 20–28. doi:10.1016/j.addma.2017.03.004.

[26] D. Yan, C.C. Tasan, D. Raabe, High resolution in situ mapping of microstrain and microstructure evolution reveals damage resistance criteria in dual phase steels, Acta Mater. 96 (2015) 399–409. doi:10.1016/j.actamat.2015.05.038.

[27] C. Cayron, ARPGE: A computer program to automatically reconstruct the parent grains from electron backscatter diffraction data, J. Appl. Crystallogr. 40 (2007) 1183–1188. doi:10.1107/S0021889807048777.

[28] A. Barnoush, H. Vehoff, Recent developments in the study of hydrogen embrittlement: Hydrogen effect on dislocation nucleation, Acta Mater. 58 (2010) 5274–5285. doi:10.1016/j.actamat.2010.05.057.

[29] M. Koyama, H. Springer, S. V. Merzlikin, K. Tsuzaki, E. Akiyama, D. Raabe, Hydrogen embrittlement associated with strain localization in a precipitation-hardened Fe-Mn-Al-C light weight austenitic steel, Int. J. Hydrogen Energy. 39 (2014) 4634–4646. doi:10.1016/j.ijhydene.2013.12.171.

[30] J. Kang, D.S. Wilkinson, J.D. Embury, M. Jain, Microscopic Strain Mapping Using Scanning Electron Microscopy Topography Image Correlation at Large Strain, J. Strain Anal. Eng. Des. 40 (2005) 559–570. doi:10.1243/030932405X16151.

[31] L. Morsdorf, O. Jeannin, D. Barbier, M. Mitsuhara, D. Raabe, C.C. Tasan, Multiple mechanisms of lath martensite plasticity, Acta Mater. 121 (2016) 202–214. doi:10.1016/j.actamat.2016.09.006.

[32] J.C. Williams, R.G. Baggerly, N.E. Paton, Deformation behavior of HCP Ti-Al alloy single crystals, in: Metall. Mater. Trans. A Phys. Metall. Mater. Sci., 2002: pp. 837–850. doi:10.1007/s11661-002-0153-y.

[33] T.B. Britton, F.P.E. Dunne, A.J. Wilkinson, On the mechanistic basis of deformation at the microscale in hexagonal close-packed metals, Proc. R. Soc. A Math. Phys. Eng. Sci. 471 (2015) 20140881. doi:10.1098/rspa.2014.0881.

[34] J. V. Bernier, J.S. Park, A.L. Pilchak, M.G. Glavicic, M.P. Miller, Measuring stress distributions in Ti-6Al-4V using synchrotron X-ray diffraction, in: Metall. Mater. Trans. A Phys. Metall. Mater. Sci., 2008: pp. 3120–3133. doi:10.1007/s11661-008-9639-6.

[35] W.Z. Abuzaid, M.D. Sangid, J.D. Carroll, H. Sehitoglu, J. Lambros, Slip transfer and plastic strain accumulation across grain boundaries in Hastelloy X, J. Mech. Phys. Solids. 60 (2012) 1201–1220. doi:10.1016/j.jmps.2012.02.001.

[36] F. Bridier, P. Villechaise, J. Mendez, Analysis of the different slip systems activated by tension in a α/β titanium alloy in relation with local crystallographic orientation, Acta Mater. 53 (2005) 555–567. doi:10.1016/j.actamat.2004.09.040.

[37] Y.H. Zhao, Y.Z. Guo, Q. Wei, A.M. Dangelewicz, C. Xu, Y.T. Zhu, T.G. Langdon, Y.Z. Zhou, E.J. Lavernia, Influence of specimen dimensions on the tensile behavior of ultrafine-grained Cu, Scr. Mater. 59 (2008) 627–630. doi:10.1016/j.scriptamat.2008.05.031.